\documentstyle[pra,epsf,aps]{revtex}

\def\comment#1{}

\begin{document}
\title{{
Spontaneous Generation of Torsion Coupling
of Electroweak Massive Gauge Bosons
}}
\author{Hagen Kleinert%
 \thanks{Email: kleinert@physik.fu-berlin.de~~~ URL:
http://www.physik.fu-berlin.de/\~{}kleinert \hfil
}}
\address{Institut f\"ur Theoretische Physik,\\
Freie Universit\"at Berlin, Arnimallee 14,
14195 Berlin, Germany}

\maketitle
\begin{abstract}
We derive the coupling to torsion
of
massive electroweak vector bosons
generated
by the Higgs mechanism.
\end{abstract}
~\\
{\bf 1.~}In the Poincar\'e gauge formulation  \cite{hehl}
of Einstein-Cartan gravity, the electromagnetic
field cannot couple minimally to torsion since
this would destroy
gauge invariance.
If
torsion does not propagate, so that the torsion field is
confined
to exist only inside of elementary particles,
this would not matter, since the propagation of photons
within matter
is much stronger modified by
electromagnetic dispersion and absorption
than by any conceivable  gravitational torsion field.
In the physically only interesting case of
a {\em propagating torsion\/}, however,
a non-gauge-invariant
coupling would have the fatal consequence
that the photon
would become massive. Since the photon mass
can be estimated experimentally
to be smaller than
$3\times 10^{-27}$ eV, this would lead to the conclusion
that the torsion
field in the universe is so small
that there is no need for contriving theories
for its possible properties, or that the photon does
not couple to torsion via the
affine covariant derivative.

Massive vector bosons, on the other hand,
such as the $ \rho $-meson,
whose wave function
has a large amplitude in a state
of a quark and an antiquark in an $s$-wave spin triplet channel,
should certainly couple to torsion via their quark
content.

By analogy with photons, the fundamental
action describing electroweak processes
should contain no minimally coupled torsion
in the gradient terms of the bare vector bosons $W$ and $Z$.
However, these particles acquire a mass via the Meissner-Higgs effect
which makes them essentially composite particles,
their fields
being a
mixture
of the original massless
vector fields and the Higgs fields.
By analogy with
the
massive $ \rho $-vector field,
we
could expect
that also the massive electroweak vector fields couple to
torsion, and the question arises how the  Meissner-Higgs effect
is capable
of  generating such a coupling.

In general, we do not know the answer to this problem.
In this note we would like
to show how such a coupling does arise
if the torsion is of the gradient type
\begin{equation}
S_{\mu\nu}{}^{\lambda} ( x ) =
 \frac{1}{2}
\left[ \delta_{\mu}^{\,\,\,\,\lambda} \partial_{\nu} \sigma ( x ) -
\delta_{\nu}^{\,\,\,\,\lambda} \partial_{\mu} \sigma ( x ) \right] \, .
\label{S2}
\end{equation}

~\\
{\bf 2.~}In the standard Poincar\'e gauge formulation  \cite{hehl} of
gravity it is immediately obvious that
 a minimally-coupled scalar Higgs field
with an
action
\begin{equation}
{\cal A}[\phi]= \int d^4x\, \sqrt{-g}
\left({\frac 1 2} g^{\mu \nu}| \nabla_\mu \phi \nabla_\nu \phi| -
\frac{m^2}{2}| \phi|^2 -\frac{ \lambda }{4}|\phi^2|^2\right)
\label{sfAF}
\label{@}\end{equation}
cannot
equip a previously uncoupled massless vector field with a torsion coupling.
 For simplicity, we consider only
a
simple Ginzburg-Landau-type theory with a
complex field to avoid inessential complications.
As usual,   $g=\det g_{\mu \nu }$
denotes the determinant of the metric $g_{\mu \nu }(x)$, and $\nabla_\mu$
is
the electromagnetic
covariant derivatives $\nabla_\mu=\partial _\mu-ieV_\mu$.
The square mass is negative, so that
the Higgs field has
a nonzero expectation value with $|\phi|^2=-m^2/ \lambda $.
From the derivative term, the vector field
acquires a mass term
$e^2|\phi|^2V^\mu V_\mu/2$, leading to the
a free part of
the vector boson action
\begin{equation}
{\cal A}[V]= \int d^4x\, \sqrt{-g}\left(- {\frac 1 4}F_{\mu \nu }
F^{\mu \nu }
-
\frac{e^2m^2}{ 2 \lambda }
V_ \nu  V^ \mu\right)
,\label{@vb}\end{equation}
where $F_{\mu \nu }$  the covariant curl
$F_{\mu \nu }\equiv
\partial _\mu V_ \nu
-\partial _\nu V_ \mu $.
Of course, the covariant curl of the
nonabelian
electroweak vector bosons
would also have self-ineractions, which can however
be ignored
in the present dicussion
since we are only interested in the free-particle propagation.

Since the Meissner-Higgs effect creates the mass
of the vector bosons by mixing the uncoupled
bare vector boson with the scalar Higgs field,
it is
obvious that
the  massive vector bosons
can couple to torsion
only if the scalar Higgs field
has such a coupling.
Indeed, it
has recently been emphasized
\cite{PI,kl}
that, contrary
to common belief \cite{geodesics},
trajectories of scalar particles
should be experience a torsion force.
This conclusion was reached
by a careful reinvestigation
of the geometric properties
of the variational procedure
of the action. Taking into account
the fact that
in the presence of torsion
parallelograms
exhibit a closure failure, the variational
procedure required a modification
of this procedure \cite{PI,fiz,pel}
which led to the conclusion
that
scalar
particles should  move
along
autoparallel trajectories
rather than geodesic ones as derived
from a minimally coupled scalar field action
\cite{geodesics}.
The modification of the variational procedure
was suggested to us
by the
close analogy
of
spaces with torsion
with crystals
containing defects \cite{GFCM}.

~\\
{\bf 3.~}So far, the classical trajectories have been
quantized consistently with
unitarity of time evolution only
for a gradient torsion (\ref{S2}),
and for a completely antisymmetric torsion
\cite{PI}.  In the case of gradient torsion
the Schr\"odinger equation turns out to be driven
by the Laplace
 operator $g^{\mu \nu }D_\mu D_ \nu $,
where $D_\mu$
is the covariant derivative involving the full affine connection
$ \Gamma _{\mu \nu }{}^ \lambda $,
including torsion.
It differs from the
Laplace-Beltrami
operator in torsion-free spaces
$ \Delta \equiv \sqrt{|g|}^{-1}\partial_\mu\sqrt{|g|}g^{\mu\nu}\partial_\nu$
by a term $-2S^{ \nu  \lambda }{}_\lambda \partial _\nu=-3(\partial ^ \nu  \sigma )\partial  _\nu $.
This operator, however,
is hermitian  only
in a scalar product
which contains  a factor $e^{-3  \sigma  }$ \cite{foot}.
In the case of totally antisymmetric torsion,
the two Laplace operators are equal and the original scalar product
ensures hermiticity and thus unitarity of  time evolution.
Such a torsion drops also out from
the classical equation of motion,
so that autoparallel and geodesic trajectories coincide.
For this reason we shall continue the discussion only
for gradient torsion.

The gradient torsion has the advantage that
it can be incorporated into the classical
action of a scalar point particle
in such a way that
the modification of the variational procedure
found in \cite{fiz,pel}
becomes superfluous. The modified
action reads
for a massive particle
 \cite{klpel}
\begin{equation}
{\cal A}[x]= -mc\int d\tau\, e^{ \sigma(x) }
\sqrt{g_{\mu \nu}(x)) \dot x^\mu \dot x^\nu}= -mc\int ds\, e^{ \sigma (x(s))},\label{pA2}
\label{classac}\end{equation}
where $\tau $ is an arbitrary parameter and $s$ the proper time.
From the Euler-Lagrange equation
we find that for $\tau =s$,
the Lagrangian under the integral
is a constant of motion, whose value  is, moreover, fixed by
the mass shell constraint
\begin{equation}
 L=	 e^{ \sigma(x) }
\sqrt{g_{\mu \nu}(x)) \dot x^\mu \dot x^\nu}\equiv 1,~~~~\tau =s.
\label{@L}\end{equation}
The necessity of
a
factor $e^{-3 \sigma (x)}$ in the scalar product
discovered in \cite{PI}
became the basis
of a series of studies in general relativity
\cite{saa,fizn}. In the latter work,
the action of a
relativistic free scalar field
$\phi$
was found to be
\begin{equation}
{\cal A}[\phi]= \int d^4x\, \sqrt{-g} e^{-3 \sigma } \,
\left({\frac 1 2} g^{\mu \nu}| \nabla_\mu \phi \nabla_\nu \phi| -
\frac{m^2}{2}| \phi|^2 e^{-2 \sigma }\right).
\label{sfAF}
\label{@freef}\end{equation}
The associated Euler-Lagrange equation is
\begin{equation}
 D_\mu D^\mu
 \phi + m^2e^{-2 \sigma(x) } \phi =0,
\label{sfEG}
\end{equation}
whose eikonal approximation $\phi(x)\approx e^{i{\cal E}(x)}$
yields
the following equation for the phase ${\cal E}(x)$  \cite{fizn}:
\begin{equation}
e^{2 \sigma(x) }g^{\mu \nu } (x)[\partial_\mu {\cal E}(x)][ \partial_ \nu{\cal E}(x)]   = m^2.
\label{@Eik}\end{equation}
Since $\partial_\mu {\cal E}$ is the momentum of the particle,
the replacement $\partial_\mu {\cal E}\rightarrow m\dot x_\mu$
shows that the eikonal equation (\ref{@Eik})
guarantees the constancy of the Lagrangian
(\ref{@L}), thus describing autoparallel trajectories.

~\\
{\bf 4.~}Apart from the factor
$e^{-3  \sigma (x) }$ accompanying the
volume integral,
the  $ \sigma $-field couples to the scalar field like a
dilaton, the power of $e^{- \sigma }$ being determined
by the dimension of the associated term.
If we therefore add to the free-field action
(\ref{@freef}) a quartic self-interaction
to have a Meissner-Higgs effect,
this self-interaction
will not carry an extrafactor  $e^{- \sigma }$, so that the
proper Higgs
action in the presence of gradient torsion reads
\begin{equation}
{\cal A}[\phi]= \int d^4x\, \sqrt{-g} e^{-3 \sigma } \,
\left({\frac 1 2} g^{\mu \nu}| \nabla_\mu \phi \nabla_\nu \phi| -
\frac{m^2}{2}| \phi|^2 e^{-2 \sigma }-\frac{ \lambda }{4}|\phi^2|^2\right)
\label{sfAF}
\label{@}\end{equation}
If $m^2$ is negative, and the torsion
depends only weakly on spacetime,
the Higgs field has a smooth vacuum expectation value
\begin{equation}
|\phi|^2=-\frac{m^2}{ \lambda }e^{-2 \sigma }.
\label{@mph}\end{equation}
The smoothness of the torsion field is required
over a length scale of
the Compton wavelength
of the Higgs particle, i.e. over a distance of the order
 $1/20$GeV $\approx 10^{-15}$ cm.
For a torsion field of gravitational origin, this smoothness
will certainly be guaranteed.
From the gradient term in (\ref{sfAF})
we then extract in the gauge
$\phi=$real the
mass term of the vector bosons
\begin{equation}
\int d^4x\, \sqrt{-g} e^{-3 \sigma }\,\frac{1}{2}m_V^2 e^{-2\sigma(x)}V^\mu V_\mu
\label{@}\end{equation}
where
\begin{equation}
m_V^2= -\frac{e^2}\lambda{m^2},~~~~m^2<0.
\label{@}\end{equation}
Taking the physical scalar product in the presence of torsion into account,
we obtain for the massive vector bosons
the free-field action
\begin{equation}
{\cal A}[V]= \int d^4x\, \sqrt{-g} e^{-3 \sigma }\,\left(- {\frac 1 4}F_{\mu \nu }
F^{\mu \nu }
+
m_V^2e^{-2\sigma(x)}
V_ \nu  V^ \mu\right)
.\label{@vb2}\end{equation}
The appearance of the
factor $e^{-2 \sigma }$
in the mass term guarantees again the same
autoparallel trajectories   in the eikonal approximation
as for spinless particles
in the action (\ref{@freef}).

Note that the scalar product factor $e^{-3 \sigma (x)}$
implies a coupling to torsion also for the
massless vector bosons which is fully compatible
with gauge invariance.
Due to the
symmetry between $Z$-boson and photon, this
factor must be present also in the electromagnetic action.

~\\
{\bf 5.~}Let us end by remarking that
autoparallel trajectories
may be considered as a manifestation
of a {\em nonholonomic mapping principle\/}
proposed in
\cite{kl,PI}
which transforms classical  equations of
motion from flat space
to spaces with curvature and torsion.
This principle was an essential tool for finding the solution
of a completely different fundamental problem,
the path integral of the hydrogen atom \cite{PI}.

Autoparallel trajectories are also the most natural trajectories
if a
 space with torsion is constructed by
a nonholonomic embedding a Riemann-Cartan space in a
flat space \cite{embed}.
The are, moreover, the only trajectories
which do not violate the
{\em universality principle\/}
of
spin and angular momentum in the coupling
of fundamental particles to torsion.
As was shown in \cite{spin},
this principle
greatly restricts
such couplings,
since the spin of  a fundamental particle
is always a fluctuating mixture of
orbital and spin angular momenta
 of its constituents.
This mixture cannot be resolved
in a representation
of the Poincar\'e group for the composite particle, since
its states are labeled  by  the quantum numbers $s,s_3$ of
the total spin.
This blindness
will
be inherent
in
any
gauge  theory
of this group, if it is to be
compatible with the physics of elementary particles,
as long as we do not possess an {\em ultimate theory\/}
of these particles,
which only string people claim to possess, in spite of a complete
disagreement with spacetime dimensions and particle spectra of
the world in which we live.

~~\\Acknowledgement:\\~\\
The author
is grateful to Prof. F.W. Hehl
for numerous interesting discussions and to
Dr. A. Pelster for useful comments.


\begin{thebibliography}{11}
\bibitem{foot}
See Section 11.4
in Ref.~\cite{PI}. Note that the normalization of the
$ \sigma $-field is normalized differently from
the present one by a factor $2/3$.
There we introduced $ \delta $ via the relation
$S_{\mu \nu }{}^ \nu =\partial _\mu \sigma $,whereas here
 $S_{\mu \nu }{}^ \nu =(3/2)\partial _\mu \sigma $.
\end{thebibliography}
\end{document}